\documentclass[prx,aps,twocolumn,showpacs,superscriptaddress,nofootinbib]{revtex4-2}

\usepackage{titlesec}

\usepackage[english]{babel}
\usepackage{bm}
\usepackage{amssymb}
\usepackage{color, soul}
\usepackage{amsmath}
\usepackage{amstext}
\usepackage{latexsym}
\usepackage{graphicx}
\usepackage{mathtools}
\usepackage[ruled]{algorithm2e}
\usepackage[dvipsnames]{xcolor}
\usepackage[colorlinks=true,citecolor=Blue,linkcolor=RubineRed,urlcolor=Blue]{hyperref}
\usepackage{tabularx}
\usepackage[capitalise,sort&compress]{cleveref}

\begin{document}

\title{Asymptotic Freedom and Finite-size Scaling of Two-dimensional Classical Heisenberg Model}

\author{Dingyun Yao}
\thanks{These two authors contributed equally to this work}
\affiliation{Hefei National Research Center for Physical Sciences at the Microscale and School of Physical Sciences, University of Science and Technology of China, Hefei 230026, China}

\author{Chao Zhang}
\thanks{These two authors contributed equally to this work}
\affiliation{Hefei National Research Center for Physical Sciences at the Microscale and School of Physical Sciences, University of Science and Technology of China, Hefei 230026, China}
\affiliation{Department of Physics, Anhui Normal University, Wuhu, Anhui 241000, China}

\author{Z. Y. Xie}
\affiliation{School of Physics, Renmin University of China, Beijing 100872, China}
\affiliation{Key Laboratory of Quantum State Construction and Manipulation (Ministry of Education), Renmin University of China, Beijing 100872, China}

\author{Zhijie Fan}
\email{zfanac@ustc.edu.cn}
\affiliation{Hefei National Laboratory, University of Science and Technology of China, Hefei 230088, China}
\affiliation{Shanghai Research Center for Quantum Science and CAS Center for Excellence in Quantum Information and Quantum Physics, University of Science and Technology of China, Shanghai 201315, China}

\author{Youjin Deng}
\email{yjdeng@ustc.edu.cn}
\affiliation{Hefei National Research Center for Physical Sciences at the Microscale and School of Physical Sciences, University of Science and Technology of China, Hefei 230026, China}
\affiliation{Hefei National Laboratory, University of Science and Technology of China, Hefei 230088, China}
\affiliation{Shanghai Research Center for Quantum Science and CAS Center for Excellence in Quantum Information and Quantum Physics, University of Science and Technology of China, Shanghai 201315, China}

\date{\today}

\begin{abstract}
The classical Heisenberg model is one of the most fundamental models in statistical and condensed matter physics. Extensive theoretical and numerical studies suggest that, in two dimensions, this model does not exhibit a finite-temperature phase transition but instead manifests asymptotic freedom. However, some research has also proposed the possibility of a Berezinskii-Kosterlitz-Thouless (BKT) phase transition over the years. In this study, we revisit the classical two-dimensional (2D) Heisenberg model through large-scale simulations with linear system sizes up to $L=16384$. Our Monte-Carlo data, without any extrapolation, clearly reveal an exponential divergence of the correlation length $\xi$ as a function of inverse temperature $\beta$, a hallmark of asymptotic freedom. Moreover, extrapolating $\xi$ to the thermodynamic limit in the low-temperature regime achieves close agreement with the three-loop perturbative calculations. We further propose a finite-size scaling (FSS) ansatz for $\xi$, demonstrating that the pseudo-critical point $\beta_L$ diverges logarithmically with $L$. The thermodynamic and finite-size scaling behaviors of the magnetic susceptibility $\chi$ are also investigated and corroborate the prediction of asymptotic freedom. Our work provides solid evidence for asymptotic freedom in the 2D Heisenberg model and advances understanding of finite-size scaling in such systems.
\end{abstract}

\maketitle

\section{Introduction} \label{sec1}
The O$(N)$ spin models are among the most fundamental and extensively studied models in statistical mechanics and condensed matter physics, having attracted widespread attention over the past century. Given a $d-$dimensional lattice, its Hamiltonian is given by
\begin{equation}
    H=-\sum_{\langle i,j\rangle}\boldsymbol{S}_i\cdot\boldsymbol{S}_j,
\end{equation}
where $\boldsymbol{S}_i$ refers to an $N$-dimensional unit spin on the $i$-th site, and the summation runs over all nearest-neighbor sites $\langle i,j\rangle$ of the lattice. For $N = 1, 2, 3$, the O$(N)$ models correspond to the well-known Ising, XY, and Heisenberg models, respectively. Studies of the O$(N)$ spin model have not only deepened our understanding of phase transitions and critical phenomena but also advanced numerous techniques in field theory, such as renormalization group (RG) methods~\cite{PhysRevB.4.3174, PhysRevB.4.3184}, $\epsilon$-expansion~\cite{PhysRevLett.28.240}, and $1/N$-expansion~\cite{PhysRev.176.718}. It is known that the O$(N)$ spin models have a lower critical dimension, below and at which spontaneous symmetry breaking is prohibited. For $N\leq1$, e.g., in the Ising model and the self-avoiding walk model (the $N\rightarrow0$ limit), the lower critical dimension is $d=1$. In contrast, for $N\geq2$ cases, the lower critical dimension is $d=2$. Thus, the O$(N)$ spin models with $N\geq2$ cannot exhibit spontaneous continuous symmetry breaking in $d\le2$, which is encapsulated by the Mermin-Wagner theorem~\cite{PhysRevLett.17.1133}. However, the absence of spontaneous symmetry breaking does not necessarily preclude finite-temperature phase transitions. The two-dimensional (2D) XY model, for instance, undergoes the celebrated Berezinskii-Kosterlitz-Thouless (BKT) topological phase transition into a low-temperature phase with quasi-long-range order (QLRO)~\cite{JMKosterlitz_1973, RevModPhys.89.040501}. For quite a while, people wondered whether the 2D Heisenberg model similarly exhibits QLRO at low temperatures like the XY model or lacks any finite-temperature phase transition altogether.

A field-theoretical perspective offers critical insights. The O$(N)$ spin model can be described by the non-linear $\sigma$ model in the continuum limit, whose partition function is
\begin{equation}
    Z=\int \mathcal{D}[\boldsymbol{s}(\boldsymbol x)]\delta(|\boldsymbol{s}(\boldsymbol x)|^2-1)\exp{[-\frac{1}{2g}\int\mathrm{d}^dx(\nabla \boldsymbol{s}(\boldsymbol{x}))^2]},
\label{action}
\end{equation}
where $\boldsymbol{s}(\boldsymbol{x})$ is a $d$-dimensional spin vector field satisfying the unit norm constraint of the O$(N)$ spins and $\frac{1}{2g}$ relates to the inverse temperature $\beta$ in the corresponding O$(N)$ spin model. At low temperatures, after an infinitesimal rescaling $\boldsymbol{x}\to\boldsymbol{x}e^{-l}$ with $0<l\ll 1$, one can obtain the RG equation at a dimension near $d=2$ in perturbation theory~\cite{10.1093/oso/9780198834625.003.0019}:
\begin{equation}
    \frac{\mathrm{d}t}{\mathrm{d}l}=\frac{N-2}{2\pi}t^2-(d-2)t+O(t^3,t^2(2-d)),
\label{RG_eq}
\end{equation}
with bare coupling $t=g\Lambda^{d-2}$, and the momentum cut-off $\Lambda=1/a$. For $d \leq 2$ and $N > 2$, the right-hand side of Eq.~\eqref{RG_eq} is always positive. As a result, $t$ flows towards infinity in the parameter space as the energy scale decreases. This indicates the absence of a fixed point at a finite temperature, and the system does not exhibit a finite-temperature phase transition. From another perspective, $t$ flows towards zero as the energy scale increases. This vanishing of the bare coupling constant $t$ at large energy scales, or equivalently at short distances, is the hallmark of asymptotic freedom. Based on this observation, Polyakov and Zinn-Justin first proposed that the 2D O$(N)$ models with $N \geq 3$ exhibit asymptotic freedom~\cite{POLYAKOV197579, PhysRevLett.36.691}. In addition, the perturbative RG computations predict the low-temperature behavior of the correlation length $\xi$. For the 2D Heisenberg model ($d=2$ and $N=3$), the one-loop perturbative calculation in Eq.~\eqref{RG_eq}, suggests an exponential divergence of $\xi$ as
\begin{equation}
    \xi=c_\xi\cdot e^{2\pi \beta},
\label{1loop}
\end{equation}
with $c_\xi$ being some constant. In the two- and three-loop perturbative calculations, $\xi$ can be respectively expressed by~\cite{CARACCIOLO1994141, FALCIONI1986671, PhysRevB.54.990}:
\begin{equation}
    \xi=c_\xi\cdot \frac{e^{2\pi \beta}}{2\pi\beta}
    \label{2loop}
\end{equation}
and
\begin{equation}
    \xi=c_\xi\cdot \frac{e^{2\pi \beta}}{2\pi\beta}(1-\frac{0.091}{\beta}+O(\beta^2)).
    \label{3loop}
\end{equation}
The constant $c_\xi$ cannot be directly calculated in perturbation theory but has been later determined as $c_\xi=\frac{e^{1-\pi/2}}{8\times2^{5/2}}$ using the Bethe ansatz~\cite{HASENFRATZ1990522, HASENFRATZ1990529}.

Although there is little doubt about the absence of finite-temperature phase transitions in the 2D non-linear $\sigma$ model in the context of the perturbation theory, its equivalence with the 2D Heisenberg model holds only in the continuum limit. Corrections arising from finite lattice spacing may introduce additional effects. Moreover, the non-perturbative validity of asymptotic freedom in the 2D non-linear $\sigma$ model has also been questioned~\cite{PATRASCIOIU1991173, PhysRevD.56.2555}. 

To verify asymptotic freedom and clarify the relationship between the discrete lattice model and the non-linear $\sigma$ model, a series of numerical studies on the 2D O$(N)$ models (with \( N \geq 3 \)) have been conducted~\cite{PhysRevD.43.2687, HASENFRATZ1990233, WOLFF1990335, WOLFF1990581,LUSCHER1991221,PhysRevLett.70.1735, PhysRevLett.75.1891, PhysRevLett.74.2969,PhysRevLett.60.2235, ALLES1997513, ALLES1997677, PhysRevD.59.067703, BALOG2003506}. However, in early Monte Carlo (MC) simulations of the 2D O$(3)$ model, due to the limitation of finite-size systems, the maximum correlation length obtained is only approximately at the order of $O(100)$, which shows a considerable deviation from the theoretical prediction~\cite{PhysRevD.43.2687, HASENFRATZ1990233, WOLFF1990335, WOLFF1990581}. Later, Lüscher $et~ al.$~\cite{LUSCHER1991221} employed a finite-size scaling (FSS) analysis to extrapolate the data to larger volumes and compute the running coupling in asymptotically free theories. Following this, Ref.~\cite{PhysRevLett.70.1735, PhysRevLett.75.1891, PhysRevLett.74.2969} similarly extrapolated the simulation data, obtaining correlation lengths up to $10^5$ at temperature as low as $\beta = 3$. Although the obtained correlation lengths still show a $4\%$ discrepancy with the three-loop result, the relationship between the extrapolated correlation length and the energy was found to satisfy the three-loop prediction near $\beta=3$~\cite{PhysRevLett.75.1891}, providing rather strong evidence for the asymptotic freedom in the 2D Heisenberg model and the 2D non-linear $\sigma$ O\((3) \) model. Additionally, a series of numerical studies also support the asymptotic freedom in the 2D O\((3) \), O\((4) \), and O\((8) \) models~\cite{PhysRevLett.60.2235, ALLES1997513, ALLES1997677, PhysRevD.59.067703, BALOG2003506}.

Although asymptotic freedom in the 2D O\((N) \) and non-linear $\sigma$ O\((N)\) models with $N\geq3$ is widely accepted, skepticism persists in some studies. The points of contention include concerns that the \( 1/N \) expansion at low temperature may deviate from the true physical behavior~\cite{PATRASCIOIU1995596}, the extrapolation methods used in Refs.~\cite{PhysRevLett.70.1735, PhysRevLett.75.1891, PhysRevLett.74.2969} is considered unreliable \cite{PhysRevLett.73.3325, PhysRevLett.76.1178}, and numerical simulations on finite systems cannot fully provide evidence for the properties of the system in the continuum limit~\cite{PATRASCIOIU1996279}. Some works argue for a finite-temperature phase transition in the 2D O$(3)$ model~\cite{PATRASCIOIU1993184, PhysRevB.54.7177, Kapikranian_2007}.

Recently, the 2D Heisenberg model has been re-examined using tensor network techniques~\cite{10.21468/SciPostPhys.11.5.098, PhysRevE.106.014104, PhysRevE.107.014117}. In Ref.~\cite{10.21468/SciPostPhys.11.5.098}, the observed correlation length divergence seems to be compatible with both the assumption of asymptotic freedom and the possibility of a finite-temperature phase transition. The authors suggested that a BKT phase transition near $T=0.509$ cannot be ruled out. Motivated by the ongoing controversy about the 2D Heisenberg model, we revisit the classical 2D Heisenberg model using large-scale MC simulations. We use a variant of the Swendsen-Wang (SW) algorithm~\cite{PhysRevLett.58.86} and perform extensive simulations of systems with linear sizes ranging from \( L = 32 \) up to \( 16384 \), containing more than $2.5\times 10^8$ lattice sites in the most significant case. Details of the simulation can be found in Sec.~\ref{sec2}.

\begin{figure}[t]
    \centering
    \includegraphics[width=\linewidth]{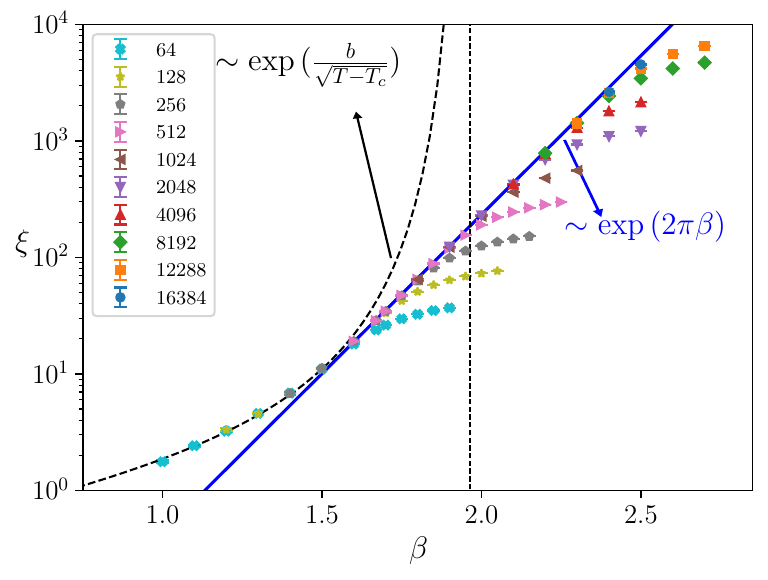}
    \caption{Exponential growth of correlation length at low temperatures. The correlation length $\xi$ is plotted as a function of $\beta$ in a semi-log coordinate for various system sizes. The growth of $\xi$ can be well described by the blue line representing $A e^{2\pi\beta}$ with $A \approx 0.0008$. The black line, representing $e^{-1.7}\cdot\exp(1.628/\sqrt{T-0.509})$, fitted by Ref.~\cite{10.21468/SciPostPhys.11.5.098}, refers to the diverging curve of $\xi$ near the possible BKT phase transition point $T_c=0.509$. This line aligns well with our data points at higher temperatures but deviates significantly as the temperature decreases.}
    \label{fig1}
\end{figure}

We first examine the temperature dependence of the correlation length $\xi$. Leveraging large-scale MC simulations, we directly measure \( \xi \) as a function of \( \beta \), as shown in Fig.~\ref{fig1}. For $\beta \le 2.2$, the behavior of measured $\xi$ faithfully captures the thermodynamic-limit behavior: at each $\beta$, $\xi$ converges to a $L$-independent value as $L \rightarrow \infty$. Even for higher $\beta$ up to $2.5$, the behavior of $\xi$ closely approximates that of the thermodynamic limit. The MC data unambiguously demonstrates an exponential divergence of $\xi$ with increasing $\beta$, and confirms the coefficient $2\pi$ in the exponent (Eq.~\eqref{1loop}), which reliably serves as a distinct signature of asymptotic freedom. To probe the growth behavior of \( \xi \) at lower temperatures, we employ the extrapolation method introduced in Ref.\cite{PhysRevLett.75.1891}. The extrapolated $\xi$ values for $\beta \ge 3.2$, as shown in Fig.~\ref{extra_pic}, quantitatively match with the three-loop perturbative calculation and strongly support the validity of perturbation theory. Moreover, the consistency between the extrapolated values and the raw data obtained directly from large-scale simulations further confirms the reliability of the extrapolation method, addressing the concerns raised in Ref.~\cite{PhysRevLett.76.1178}.

At lower temperatures, $\xi$ becomes comparable to the finite system size $L$, and the system enters the ``FSS regime", where the thermodynamic behavior of $\xi$ in Eq.~\eqref{3loop} is no longer valid. To describe $\xi$ in this regime, we then investigate its FSS behavior. Based on the exponential divergence of \( \xi \) with $\beta$, we conjecture and confirm an FSS ansatz (see Eq.~\eqref{xi_FSS_form} and Fig.~\ref{fig2}). The formula can describe the behavior of $\xi$ in the FSS regime and is consistent with its thermodynamic behavior when $\xi\ll L$. The crossover between the two behaviors occurs at the pseudo-critical point $\beta=\beta_L$ where $\xi/L$ is at a constant of order $O(1)$. Upon our FSS ansatz of $\xi$, we show that $\beta_L$ diverges logarithmically with system size \(L\), with additional correction terms obtained from two-loop perturbation results.

We further investigate the behavior of magnetic susceptibility $\chi$ and its relationship to $\xi$, which provides critical insights into the phase transition properties of the system. During the debate surrounding the 2D O$(3)$ non-linear $\sigma$ model, Patrascioiu et al. \cite{PhysRevB.54.7177} observed from numerical data that the relationship between $\chi$ and $\xi$ satisfies the characteristics of a BKT phase transition, contradicting the asymptotic freedom predicted by the perturbation theory. Later, Allés et al. \cite{ALLES1997677} conducted numerical simulations using the tree-level improved Symanzik action \cite{SYMANZIK1983205}, arguing that the relationship between \( \chi \) and \( \xi \) neither supports a BKT phase transition nor aligns with the perturbation theory. In Sec. \ref{sec4}, we revisit this contradiction. Our results, as presented in Fig.~\ref{chi_corr}, clearly support the prediction of perturbation theory and exclude BKT-like scaling. 

Finally, we discuss the behavior of the specific heat \( C \) to demonstrate the system's thermal response as a function of temperature. Notably, the specific heat peak occurs around half the value of the phase transition temperature predicted by the mean-field theory. Moreover, the field-theoretic description also seems to become valid below this temperature; namely, after the specific heat peak, the correlation length $\xi$ enters the exponentially growing regime governed by asymptotic freedom.

The remainder of this paper is organized as follows. Sec.~\ref{sec2} briefly describes the algorithm and introduces the sampled quantities and observables. Sec.~\ref{sec3} presents the results of the correlation length $\xi$, including the numerical verification of the perturbation theory predictions and the FSS analysis of $\xi$. Sec.~\ref{sec4} shows the behaviors of the magnetic susceptibility $\chi$ and the specific heat $C$. Finally, a brief summary is given in Sec.~\ref{sec5}.

\section{Algorithm and Observables} \label{sec2}

In this work, we simulate the system using a variant of the Swendsen-Wang (SW) algorithm \cite{PhysRevLett.58.86}. The algorithm can significantly reduce the divergence of the autocorrelation time $\tau$ with increasing system size, thereby addressing the issue of critical slowing down. In 1989, a rigorous lower bound of $\tau$ under the SW algorithm was established as~\cite{PhysRevLett.63.827}:
\begin{equation}
    \tau\geq \text{Const}\times C,
\end{equation}
where $C$ is the specific heat. For the 2D Heisenberg model, the specific heat does not exhibit any divergence at varying temperatures. Thus, it satisfies the precondition for the SW algorithm to be effective.
In our simulations, we find the autocorrelation time of the total energy (defined later) stays at a constant $\sim 5$ for system sizes ranging from 32 to 16384 at various temperatures.

The original SW algorithm was established based on the Potts model with discrete symmetry \cite{PhysRevLett.58.86}. To extend its applicability to the Heisenberg model with continuous symmetry, one map the Heisenberg model into an effective Ising representation during each MC configuration update. At each MC step, a randomly selected unit vector $\boldsymbol{S}_{\rm{ref}}$ in the three-dimensional spin space serves as a reference axis. Each Heisenberg spin $\boldsymbol{S}$ on the lattice is subsequently projected onto this reference vector, denoted as $S^{\parallel} = \boldsymbol{S}\cdot\boldsymbol{S}_{\rm{ref}}$. Only this projection component can undergo spin flipping during the update process, while the remaining orthogonal components remain invariant. This methodology effectively reduces the update mechanism for the Heisenberg system to an Ising-type update scheme. One then activates bonds between each nearest-neighbor pair of sites to form clusters according to the probability
\begin{equation}    
    p_{ij} = \left[1 - e^{-2 \beta S^{\parallel}_iS^{\parallel}_j}\right]^{+},
\end{equation}
with $[x]^+ = {\rm max}(0,x)$. The only difference with the Ising model is that the activation probability is no longer constant but depends on the specific spin configuration of the neighboring pair. After the clusters have been formed, the spins of each cluster are kept unchanged or flipped to $\boldsymbol{S}'=\boldsymbol{S}-2S^{\parallel}\boldsymbol{S}_{\rm{ref}}$ independently with equal probability of $1/2$. After all clusters are examined, each spin on the lattice is updated once; this complete update cycle is termed a \textit{sweep}.

In this work, we perform simulations for system sizes ranging from $L=32$ to $16384$ with periodic boundary conditions. For $L$ smaller than 4096, each simulation consists of \( 2 \times 10^6 \) sweeps, which yield approximately $4\times10^5$ independent samples. Systems with larger size are simulated for certain temperatures, with more than \( 2 \times 10^4 \) sweeps, yielding at least $4\times10^3$ independent samples. The resulting correlation length reaches \( 2 \times 10^4 \) without any extrapolation.

In the simulation, for a given configuration, we directly sample the following quantities:
\begin{itemize}
\item The total energy: $E=-\sum_{\langle i,j\rangle}\boldsymbol{S}_i\cdot\boldsymbol{S}_j$.
\item The magnetization density: $M=\frac{1}{L^2}|\sum_i\boldsymbol{S}_i|$.
\item The spin Fourier transformation module: $M_k=\frac{1}{L^2}|\sum_i\boldsymbol{S}_ie^{i\boldsymbol{k}\cdot\boldsymbol{r}_i}|$, where $\bm{k}=\frac{2\pi}{L}\hat{\bm{x}}$, the smallest wave vector in Fourier space.
\end{itemize}
By taking the ensemble average $\langle\cdot\rangle$ along the Markov chain, we calculate the following observables:
\begin{itemize}
\item The second moment correlation length:
    \begin{align}
    \xi_{2\text{nd}} = \frac{1}{2 \sin(|\bm{k}|/2)} \sqrt{\langle M^2 \rangle/\langle M_k^2 \rangle - 1}.
\label{Eqn2}
\end{align}
Note that the second moment correlation length $\xi_{2\text{nd}}$, obtained from the Fourier spectrum of the correlation function, is different from the exponential correlation length $\xi_{\text{exp}}$, extracted by fitting an exponential decaying correlation function. However, the two are equivalent in the disordered phase and at the phase transition point. For the 2D Heisenberg model, the ratio of these two lengths can be expressed analytically in the large-$N$ limit as \cite{FLYVBJERG1991714, BISCARI1990225}:
\begin{equation}
    \xi_{\text{2nd}}/\xi_{\text{exp}}=1-0.003225/N+O(1/N^2)
\end{equation}
The two correlation lengths only differ by a tiny factor; for $N=3$, it's about $0.1\%$, which is within the statistical noise in the MC simulation. Since the second moment correlation length can be extracted more easily through Eq.~\eqref{Eqn2}, we use it to study the correlation length of the system. For simplicity, the symbol \( \xi \) used in the following sections of this paper refers to the second moment correlation length \( \xi_{\text{2nd}} \), and the exponential correlation length is denoted explicitly as \( \xi_{\text{exp}} \).
\item The magnetic susceptibility: $\chi=L^2\langle M^2\rangle$.
\item The specific heat: $C=\beta^2(\langle E^2\rangle-\langle E\rangle^2)/L^2$.
\end{itemize}
The error bars are estimated using the conventional binning method for directly sampled quantities. For the composite observables, we employ the Jackknife method to account for the error propagation of correlated data properly. In this work, the error bars of data without extrapolation are of the size of the markers or smaller, demonstrating an excellent convergence of the MC simulation.

\section{Correlation Length}\label{sec3}
This section comprehensively analyzes the temperature-dependent evolution of the correlation length $\xi$. Figure~\ref{fig1} presents $\xi$ as a function of inverse temperature $\beta$ on a semi-logarithmic scale. For a given system size $L$, $\xi$ undergoes three distinct regimes as $\beta$ increases: the ``initial microscopic regime," the ``intermediate thermodynamic regime," and the ``FSS regime." Initially, at small values of \(\beta\), the system is in the initial microscopic regime, where \(\xi\) is comparable or not much larger than the lattice constant $a$. In this regime, finite lattice spacing plays an important role, and the field-theoretical description in Eq.~\eqref{action} does not hold. As the temperature decreases, the system enters the intermediate thermodynamic regime where \(a \ll \xi \ll L\). In this regime, the correlation length follows the field-theoretical predictions. Its behavior is $L$-independent and consistent with that of the thermodynamic limit at $L\to \infty$. As depicted in Fig.~\ref{fig1}, the data points converge for \(\beta \leq 2.2\) as $L$ grows larger, and the approximate convergence persists up to $\beta=2.5$. The agreement between the blue line and converged data points demonstrates $\xi$'s exponential dependence on $\beta$:
\begin{equation}
\xi \sim e^{2 \pi \beta},
\label{xidiverge}
\end{equation} 
consistent with the leading-order perturbative calculations in Eq.~\eqref{1loop}.
As the temperature further decreases, the system enters the FSS regime manifested as the flattening data points at large \(\beta\) in Fig.~\ref{fig1}. In this regime, $\xi$ is comparable to $L$, causing the thermodynamic-limit behavior to break down, and the system can be described by some FSS ansatz. The crossover between the intermediate thermodynamic and FSS regime occurs when $\xi/L$ is $O(1)$.

Figure.~\ref{fig1} directly demonstrates the exponential divergence described by Eq.~\eqref{xidiverge} of \(\xi(\beta)\) without any extrapolation. This clearly rules out the possibility of a BKT phase transition occurring at \(T = 0.509\) suggested in Ref.~\cite{10.21468/SciPostPhys.11.5.098}. The black line in Fig.~\ref{fig1}, taken from the fitting results in Ref.~\cite{10.21468/SciPostPhys.11.5.098}, illustrates the correlation length growth associated with a BKT phase transition at \(T = 0.509\). This curve matches our data points only when \(\xi\) is small, primarily in the initial microscopic regime. However, as $\xi$ further increases, its growth behavior gradually deviates from the seemingly BKT-like prediction. Therefore, we conclude that the putative BKT phase transition proposed in Ref.~\cite{10.21468/SciPostPhys.11.5.098}  likely arises from a relatively small angular momentum cutoff, which restricts the effective correlation length at low temperatures.

In the rest of this section, we study the behavior of $\xi$ in the intermediate thermodynamic and FSS regimes, respectively, in Sec.~\ref{sec3a} and \ref{sec3b}.
In Sec.~\ref{sec3a}, we employ the extrapolation technique to analyze the thermodynamic-limit behavior of $\xi$ at lower temperatures, and further verify its consistency with the three-loop perturbative calculations in Eq.~\eqref{3loop}. In Sec.~\ref{sec3b}, we investigate the FSS behavior of \(\xi\) and demonstrate that the pseudo-critical point $\beta_L$, characterizing the crossover between the intermediate thermodynamic regime and the FSS regime, diverges logarithmically with system size $L$.

\subsection{Asymptotic Freedom} \label{sec3a}
\begin{figure}
    \centering
    \includegraphics[width=\linewidth]{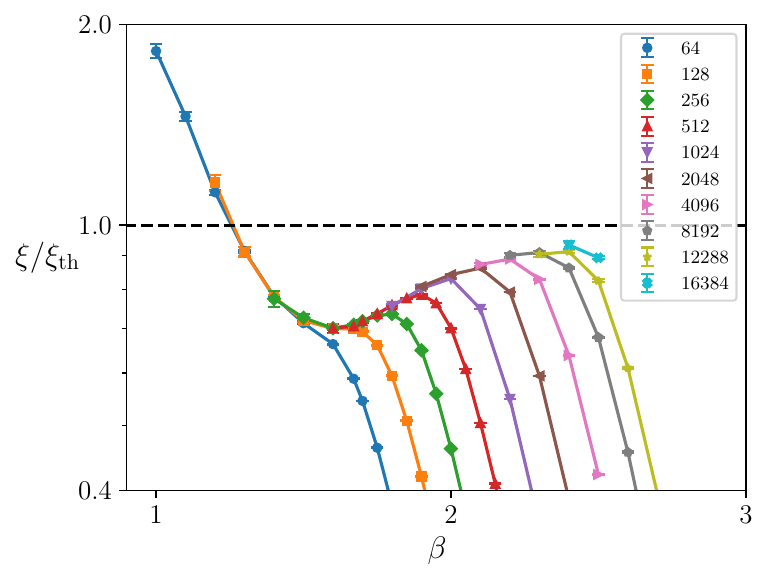}
    \caption{The ratio between the numerical value $\xi$ and the three-loop theoretical result in Eq.~\eqref{3loop}. The ratio shows a strong deviation from $1$ initially. However, as $\beta$ increases, it gradually approaches $1$, suggesting that the two may align with each other at lower temperatures.}
    \label{contrast}
\end{figure}

Through large-scale MC simulations, we have established that the correlation length $\xi$ exhibits an exponential divergence with $\beta$ as given in Eq.~\eqref{xidiverge}. This corresponds to the leading term in the three-loop theoretical prediction of Eq.~\eqref{3loop}. To further investigate whether our numerical results $\xi$ align with the full three-loop theoretical prediction $\xi_{\text{th}}$, we plot the ratio $\xi/\xi_\text{th}$ as a function of $\beta$ in Fig.~\ref{contrast}. While a distinct deviation from the three-loop result is observed in the high-temperature regime, the ratio gradually approaches $1$ as the temperature decreases. These observations strongly suggest that the perturbative theoretical predictions only set in at sufficiently low temperatures. Thus, it is necessary to measure $\xi$ at lower temperatures to verify the three-loop result. One needs to simulate larger system sizes at lower temperatures to suppress the finite-size effect. However, the brute-force simulation approach is numerically challenging and impractical. Instead, one can exploit the FSS behavior of \(\xi\) and extrapolate \(\xi\) to obtain its thermodynamic value \(\xi_\infty\) at lower temperature. In the following, we introduce the extrapolation method in detail and present our results.

The extrapolation of \(\xi\) has been applied effectively and successfully in previous Monte Carlo (MC) studies \cite{PhysRevLett.70.1735, PhysRevLett.75.1891, PhysRevLett.74.2969}. This extrapolation method is based on an FSS ansatz \cite{PhysRevLett.75.1891, PhysRevLett.74.2969}:
\begin{equation}
    \frac{\mathcal{O}(\beta,sL)}{\mathcal{O}(\beta,~L)}=F_\mathcal{O}\left(\frac{\xi(\beta,L)}{L};s\right)+O(\xi^{-\omega},L^{-\omega})
    \label{extrapolation_eq}
\end{equation}
where $\mathcal{O}$ refers to any observables affected by finite-size effect; $s$ is a fixed rescaling factor, typically set at $2$ for the convenience of MC simulations; $F_\mathcal{O}$ is a universal function of $\xi/L$, depending on the observable $\mathcal{O}$ and the rescaling factor $s$; the second term $O(\xi^{-\omega}, L^{-\omega})$ contains further correction terms, which vanishes when $\xi$ and $L$ becomes large enough. Therefore, once the form of the universal function $F_\xi$ is obtained, we can extrapolate the value of $\xi$ in a system of finite size $L$ using Eq.~\eqref{extrapolation_eq}. Specifically, by denoting $x_n=\xi(\beta,s^nL)/s^nL$ and $y_n=F_\xi(x_n,s)$, Eq.~\eqref{extrapolation_eq} can be rewritten as:
\begin{equation}
    x_{n+1}=F_\xi(x_n,s)\cdot \frac{x_n}{s}=y_n\cdot \frac{x_n}s.
    \label{iterate}
\end{equation}

By iteratively applying Eq.~\eqref{iterate}, one has a flow of $(x,y)$: $(x_0,y_0)\to(x_1,y_1)\to\cdots \to(x_n,y_n)\to\cdots$, and there are two fixed points. The first is $(x_c=\xi(\beta_c,L)/L,s)$, since at the critical point $\beta=\beta_c$, the dimensionless ratio $x_c=\xi(\beta_c,L)/L$ is a universal constant and hence $\frac{\xi(\beta_c,sL)}{\xi(\beta_c,L)}=\frac{\xi(\beta_c,sL)/sL}{\xi(\beta_c,L)/L}s=s$. The second fixed point is at \( (0,1) \), corresponding to temperatures above the critical temperature. At these temperatures, \( \xi_\infty \) is finite. When the number of iteration $n$ becomes sufficiently large, \( s^nL \gg \xi_\infty \), and $\xi(\beta,s^nL)$ converges to its thermodynamic-limit value $\xi_\infty$. In this case, \( x_n \approx \xi_\infty / s^nL \approx 0 \), and \(y_n= F_\xi(x_n,s)=\xi(\beta, s^nL) / \xi(\beta, s^{n+1}L) \approx 1 \). When iterating Eq.~\eqref{iterate} at any temperature above the critical point, the pair \( (x, y) \) will eventually flow towards the stable fixed point \( (0,1) \), which is also the process that occurs during our extrapolation. After \( n \) iterations, the correlation length at a system size of \( s^nL \) can be expressed as \( \xi_n = x_n\cdot s^nL \). Since the system size in the extrapolating process grows exponentially with $n$, only a few iterations are required for $\xi_n$ to converge to its value in the thermodynamic limit, $\xi_\infty$. For other observables $\mathcal{O}$, by iteratively applying Eq.~\eqref{extrapolation_eq} with universal functions $F_\mathcal{O}$ and $F_\xi$, their values at larger system sizes can also be extrapolated. This work only exhibits extrapolation to the correlation length $\xi$. The key question is how to determine the form of the universal function $F_\mathcal{\xi}$.

\begin{figure}
    \centering
    \includegraphics[width=\linewidth]{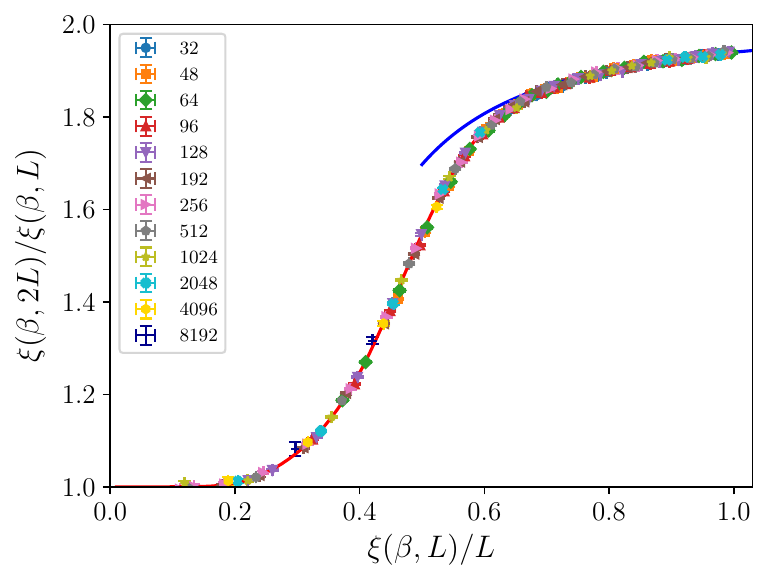}
    \caption{Verification of the FSS ansatz~\eqref{extrapolation_eq}. The vertical axis is the correlation-length ratio, $\xi(\beta,2L)/\xi(\beta, L)$, of two systems with size $2L$ and $L$, and the horizontal axis is the dimensionless ratio $\xi(\beta, L)/L$. Data points of various system sizes collapse well onto a single red curve $F_\xi(x)$, which is obtained through the fitting of Eq.~\eqref{extra_fit}. The blue line refers to the perturbation result in Eq.~\eqref{perturbative result} for large $x$.}
    \label{collapse}
\end{figure}

Before delving into the specifics of the extrapolation procedure, it is necessary first to discuss the calculation of the error bar of the extrapolated value. For a general observable $\mathcal{O}$, the error bar of the extrapolated value based on Eq.~\eqref{extrapolation_eq} comes from three primary sources: (i) The error from the observable $\mathcal{O}(\beta, L)$ before extrapolation; (ii) The error from the correlation length $\xi$ before extrapolation; (iii) The deviation of the universal function $F_\mathcal{O}$.
The first two sources stem from the statistical errors of raw data. Given a specific form of $F_\mathcal{O}$ and the extrapolation procedure in Eq.~\eqref{extrapolation_eq}, the errors in the extrapolated values can be obtained by sampling the input data through the Monte Carlo method. 
The third source refers to the systematic error from the extrapolation process itself, which arises from the uncertainty in the universal function $F_\mathcal{O}$. To estimate this contribution, different fits for $F_\mathcal{O}$ have to be attempted and the extrapolated results from each fit must be combined to obtain an estimate of the systematic error.
In practice, the contribution of the systematic error is usually smaller than that of the statistical error. A more detailed discussion can be found in Appendix~\ref{Apen_a}.

\begin{figure*}
    \centering
    \includegraphics[width=\linewidth]{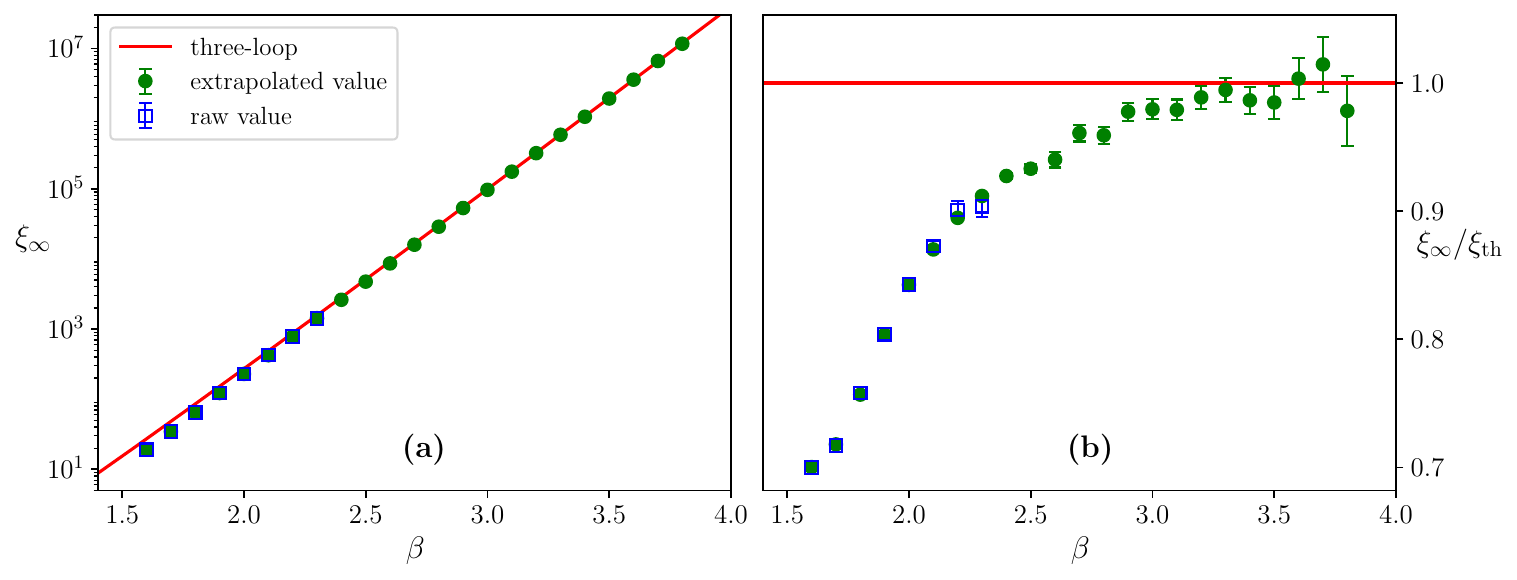}
    \caption{(a) The extrapolated thermodynamic value of the correlation length $\xi_\infty$ as a function of $\beta$. The blue squares represent the values obtained directly from the simulation at sufficiently large systems, which can reflect the thermodynamic limit. The green dots represent further extrapolations of $\xi$. The red line refers to the three-loop perturbative result in Eq.~\eqref{3loop}. (b) The ratio of data with the three-loop perturbative result versus $\beta$, demonstrating an excellent agreement for large $\beta$. }
    \label{extra_pic}
\end{figure*}

The fitting equation of $F_\mathcal{\xi}$, as in Ref.~\cite{PhysRevLett.75.1891, PhysRevLett.74.2969}, is typically chosen as the form:
\begin{equation}
    F_\mathcal{\xi}(x)=1+a_1e^{-1/x}+a_2e^{-2/x}+\cdots+a_ne^{-n/x}
\label{extra_fit}
\end{equation}
with $n$ usually no more than 12. By appropriately selecting fitting data, minimum system size, and $n$ values, we obtain a series of fitting results for $F_\xi$. Since systematic errors arising from different fits of $F_\xi$ are relatively small, for simplicity, the results presented hereafter exclusively employ the optimal fit. Details are presented in Appendix~\ref{Apen_a}.
Figure~\ref{collapse} displays $\xi(\beta,2L)/\xi(\beta,L)$ as a function of $\xi(\beta,L)/L$, together with the fitted $F_\xi(x)$. As Eq.~\eqref{extrapolation_eq} implies, the data points for different temperatures and system sizes collapse well onto the fitted $F_\xi(x)$ curve. Additionally, the curve also satisfies the asymptotic behavior predicted by the perturbation theory \cite{PhysRevLett.75.1891}:
\begin{equation}
\begin{aligned}
    F_\xi(x;s)=&s\bigg[ 1- \frac{\ln s}{8\pi}x^{-2}\\
    &-\frac14\left( \frac{\ln s}{8\pi^2}+\frac{\ln^2s}{32\pi^2}\right)x^{-4}+O(x^{-6})\bigg],
\end{aligned}
\label{perturbative result}
\end{equation}
near $x=1$. These observations confirm the reliability of both Eq.~\eqref{extrapolation_eq} and the fit of $F_\xi$.

The final extrapolated result of $\xi$, along with its deviation from the three-loop prediction in Eq.~\eqref{3loop}, is summarized in Fig.~\ref{extra_pic}. In subplot (a), we directly plot the extrapolated values of \(\xi\), signaled by green dots, as a function of \(\beta\), with the largest extrapolated value reaching \(10^7\). They seem close to the three-loop results. The blue squares represent raw data not extrapolated for smaller values of \(\beta\). Their agreement with the extrapolated data proves the reliability of the extrapolation method. In subplot (b), we plot the ratio between the extrapolated results and the three-loop perturbative prediction for further comparison. It can be observed that as $\beta$ increases, the ratio gradually converges toward 1. For $\beta\geq3.2$, the ratio almost reaches 1 within its error bar. Noting the non-smooth variation of data points, error bars of the ratio are likely to be underestimated. Furthermore, since the extrapolated results shown in Fig.~\ref{extra_pic} leverage only the optimal fit of $F_\xi$, the systematic error caused by different fits of $F_\xi$ has not been considered. Therefore, it is fair to say that, for \(\beta \geq 3.2\), the extrapolated results align with the three-loop prediction within the error bar. This strongly support asymptotic freedom in the 2D O(3) non-linear $\sigma$ model and the 2D Heisenberg model.

\subsection{Finite-Size Scaling} \label{sec3b}

In Fig.~\ref{fig1}, the data points of a given system size $L$ gradually flatten at low temperatures, indicating that the system enters the FSS regime, where the correlation length is truncated by $L$. In the following, we investigate the FSS behavior of $\xi$ in this regime. 

Before the system enters the FSS regime, The correlation length $\xi$ is solely a function of temperature: $\xi=\xi(\beta)$. However, in the FSS region, $\xi$ depends on both the temperature and the system size, which leads to an FSS ansatz,
\begin{equation}
    \xi(\beta,L)=L\tilde{\xi}(x)
\label{FSS_form}
\end{equation}
where $x$ is a function of $\beta$ and $L$, acting as a rescaled thermal scaling field in the language of RG, and $\tilde \xi(x)$ is an analytical function. Near a second-order phase transition, for instance, the rescaled thermal scaling field $x$ can be expressed as $x \equiv tL^{y_t}$, where $t = (T-T_c)/T_c$ is the reduced temperature, and $y_t$ is the thermal renormalization exponent. Thus, the system enters the FSS regime at $t=t_L\sim O(L^{-y_t})$ where the combined variable $x$ and the dimensionless ratio $\xi/L$ have the value of order $O(1)$. In this regime, the ansatz of Eq.~\eqref{FSS_form} should hold. Beyond the FSS regime when $x\gg1$, a crossover finite-size critical window can be defined by simultaneously requesting $t \to 0$ and $x \gg1$~\cite{li2024crossoverfinitesizescalingtheory}. In this crossover finite-size window, $\xi$ exhibits its thermodynamic critical behavior, $\xi \sim |t|^{-\nu}$ with $\nu$ being the correlation-length exponent, with which the FSS ansatz~\eqref{FSS_form} in the thermodynamic limit should be consistent. Starting from Eq.~\eqref{FSS_form}, to recover the thermodynamic scaling $\xi \sim |t|^{-\nu}$, it is required that: 
\begin{equation}
    \begin{aligned}
        \tilde{\xi}(x)&\to x^{-\nu}  ,x\gg1\\
        y_t&=\frac1\nu.
    \end{aligned}
\end{equation}
This ensures the consistency between the FSS ansatz and the divergence of $\xi$ in the thermodynamic limit.

\begin{figure*}[t]
    \centering
    \includegraphics[width=\linewidth]{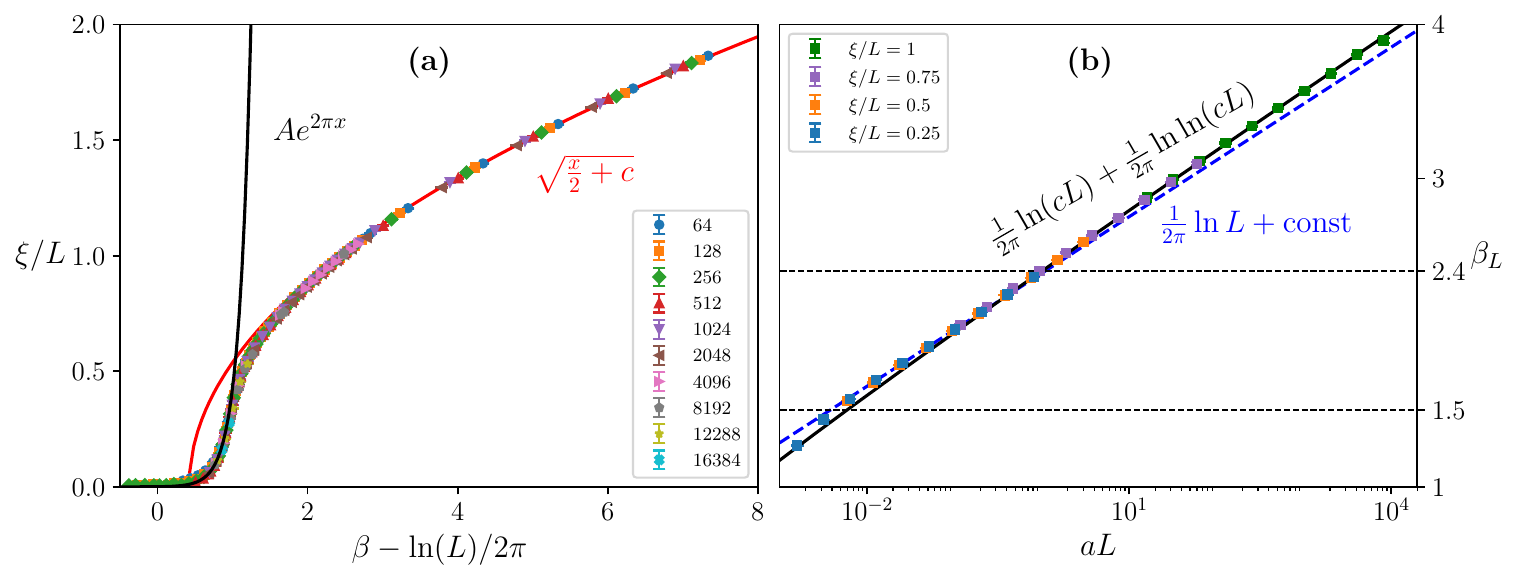}
    \caption{(a) Finite-size scaling of the ratio $\xi/L$ as a function of $x=\beta-\ln(L)/2\pi$ with no free parameters involved. The main plot shows that the data points collapse on a single curve $\Tilde{\xi}(x)$ for various system sizes ranging from $L=64$ to $16384$. The black and red lines refer to the asymptotic behaviors in Eq.~\eqref{xi_asym} and \eqref{xi_asym2} respectively, with $A=0.0008$, $c=-0.21$, and they both align well with the data points. (b) The pseudo-critical point $\beta_L$ as a function of $aL$, with $\xi/L$ fixed at different constants. The ratio $\xi/L$ is fixed at $1,0.75,0.5,0.25$ with $a=1,0.0074,0.00037,0.0001$ respectively. The data points for various system sizes ranging from $L=16$ to $8192$ are included. The blue and black lines refer to the zero- and first-order results in Eq.~\eqref{xiL_correction} respectively, with $c=2.8\times10^5$. It can be seen that the $L$-dependent behavior of $\beta_L$ has approximately three ranges: at high temperature with $\beta_L \lesssim 1.5$ it significantly deviates from the field theoretical prediction,  in intermediate temperature in range $(1.5,2.4)$ it approximately follows a logarithmic growth, and at low temperature with $\beta_L >2.4$ the subleading log-log correction appears. This interesting feature is similar to that in Fig.~\ref{fig1}.}
    \label{fig2}
\end{figure*}

Now, let us address the FSS ansatz for the 2D Heisenberg model. Similarly, we hypothesize that the FSS ansatz in Eq.~\eqref{FSS_form} holds in the FSS regime, and can smoothly transition to $\xi$'s thermodynamic-limit behavior in the crossover finite-size window, which is now defined by simultaneously requesting the inverse temperature $\beta\to\infty$ and $\xi(\beta, L)\ll L$. In the crossover finite-size window, the correlation length $\xi$ diverges exponentially with $\beta$, as predicted by the leading order of perturbative calculation in Eq.~\eqref{1loop} and observed numerically in Eq.~\eqref{xidiverge}. Therefore, to ensure the consistency between the FSS ansatz \eqref{FSS_form} and $\xi$'s exponential divergence with $\beta$ in Eq.~\eqref{1loop}, the following relation must hold:
\begin{equation}
    \tilde\xi(x)\sim\frac{e^{2\pi \beta}}{L}~~~~, \beta\to\infty, L\gg \xi.
    \label{transition_request}
\end{equation}
For simplicity, we take $x=\beta-\frac1{2\pi}\ln L$, and conjecture that within the FSS regime, $\xi(\beta,L)$ can be well described by:
\begin{equation}
    \xi(\beta,L)=L\tilde{\xi}(\beta-\frac1{2\pi}\ln L).
\label{xi_FSS_form}
\end{equation}
From Eq.~\eqref{transition_request}, $\tilde{\xi}(x)$ automatically has an asymptotic behavior of:
\begin{equation}
    \tilde\xi(x)\sim e^{2\pi x}~~~~, \beta\to\infty, -x\gg1.
\label{xi_asym}
\end{equation}
Combining with Eq.~\eqref{perturbative result}, we can also obtain the asymptotic behavior of $\tilde\xi(x)$ for $x\gg1$. Specifically, given the conjecture of Eq.~\eqref{xi_FSS_form}, $\tilde\xi(x)$ and $F_\xi(x;s)$ have the relation below:
\begin{equation}
\begin{aligned}
    F_\xi\left(\tilde{\xi}(\beta-\frac{\ln L}{2\pi});s\right)&=F_\xi\left(\frac{\xi(\beta,L)}L;s\right)\\
    &=\frac{\xi(\beta,sL)}{\xi(\beta,L)}=\frac{\tilde\xi(\beta-\frac{\ln(sL)}{2\pi})}{\tilde\xi(\beta-\frac{\ln(L)}{2\pi})}s.
\end{aligned}    
\end{equation}
Let \( x = \beta - \frac{\ln L}{2\pi} \), and substituting Eq.~\eqref{perturbative result} into the above expression, we obtain:
\begin{equation}
    \frac{\Tilde\xi(x-\frac{\ln s}{2\pi})}{\Tilde\xi(x)}=1-\frac{\ln s}{8\pi}\Tilde\xi(x)^{-2}+O(\Tilde\xi(x)^{-4}).
\end{equation}
Let $s\to1$, and denote $-\frac{\ln s}{2\pi}$ as $\mathrm{d}x$. Then, the above expression can be rewritten as:
\begin{equation}
    \frac{\tilde\xi(x+\mathrm dx)}{\tilde\xi(x)}=1+\frac{1}{4}\tilde\xi(x)^{-2}\mathrm dx+O(\tilde\xi(x)^{-4}).
\end{equation}
Thus, we obtain a differential equation of $\tilde\xi(x)$:
\begin{equation}
    \mathrm d\tilde\xi=\frac{1}{4}\tilde\xi^{-1}\mathrm dx+O(\tilde\xi^{-3}).
\end{equation}
Neglecting higher-order terms, we obtain the asymptotic behavior of $\tilde\xi(x)$ for $x\gg1$:
\begin{equation}
    \tilde\xi(x)=\sqrt{\frac12x+c}~~~~,x\gg1,
    \label{xi_asym2}
\end{equation}
where $c$ is some constant. For any given system size \( L \), as the temperature approaches zero, i.e., \( \beta \to \infty \) and $x=\beta-\frac{\ln L}{2\pi}\to\infty$, the dimensionless ratio $\xi/L$ diverges to infinity in the way described by Eq.~\eqref{xi_asym2}. Suppose we regard zero temperature as the critical point of the 2D Heisenberg model. It differs from a typical second-order or BKT phase transition, where \( \xi/L \) is a universal constant at the critical point.

To support our conjecture in Eq.~\eqref{xi_FSS_form}, we plot \(\xi/L\) as a function of \(\beta-\ln L/2\pi\) in Fig.~\ref{fig2}. The data points from different system sizes collapse nicely onto a single curve. Furthermore, the two asymptotic behaviors in Eq.~\eqref{xi_asym} and \eqref{xi_asym2} are also well satisfied. At small values of $x=\beta-\frac1{2\pi}\ln L$, the data points show nice agreement with the black curve corresponding to Eq.~\eqref{xi_asym}. On the other hand, the asymptotic behavior for $x\gg1$ in Eq.~\eqref{xi_asym2}, as denoted by the red curve with $c=-0.21$, is also well exhibited by the data points for $x>3$. This evidence strongly supports the proposed FSS formula in Eq.~\eqref{xi_FSS_form}.

Based on Eq.~\eqref{xi_FSS_form}, we can further investigate the relationship between the system's pseudo-critical point $\beta_L$ and the system size $L$. As the temperature decreases, the correlation length \(\xi\) increases. When \(\xi\) reaches the order of \(L\), it becomes truncated by the finite size of the system, and the system enters the FSS regime. This temperature is referred to as the pseudo-critical temperature $T_L$, and for convenience, we call its inverse $\beta_L=1/T_L$ as the pseudo-critical point. At this point, \(\xi/L\) is of the order $O(1)$. 
If a finite-temperature phase transition exists, the pseudo-critical point should gradually converge to the true critical point as the system size increases. This behavior provides a way to extrapolate the critical temperature and further analyze the nature of the phase transition. For 2D Heisenberg model according to Eq.~\eqref{xi_FSS_form}, by fixing \(\xi/L\) at a constant of \(O(1)\), the pseudo-critical point \(\beta_L\) should logarithmically diverge to infinity as \(L\) increases:
\begin{equation}
    \beta_L=\mathrm{Const}+\frac{1}{2\pi}\ln L.
\label{beta_L_diverge}
\end{equation}

We fix $\xi/L$ at $0.25,0.5,0.75,1$ and respectively plot the corresponding $\beta_L$ as a function of the rescaled system size $aL$. The factor $a$ takes different values for each $\xi/L$ so that the data points can collapse onto a single curve. As shown in Fig.~\ref{fig2} (b), regardless of the value of $\xi/L$ (which remains of order $O(1)$), the pseudo-critical point $\beta_L$ indeed exhibits a logarithmic divergence with $L$ for $\beta_L > 1.5$, which is consistent with Eq.~\eqref{beta_L_diverge}. Yet, at sufficiently low temperatures, $\beta_L>2.4$, the growth of data points becomes slightly faster, which is likely attributed to additional corrections originating from higher-order perturbation calculations. Specifically, the FSS form of \(\xi\) in Eq.~\eqref{xi_FSS_form} leverages the leading diverging term in Eq.~\eqref{1loop}, while the sub-leading terms due to higher-order perturbation calculations are neglected.

\begin{figure*}
    \centering
    \includegraphics[width=\linewidth]{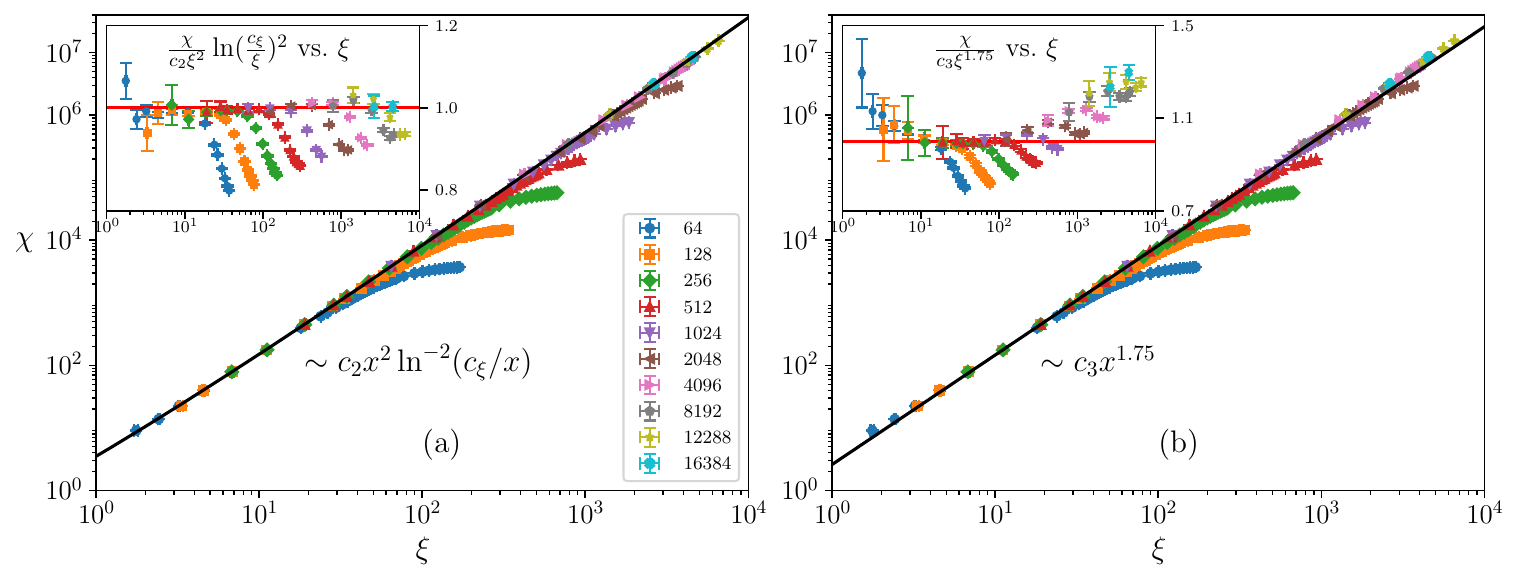}
    \caption{The comparison of the fit of Eq.~\eqref{chi_corr_relation} and \eqref{BKT_relation} with the data points. (a) The magnetic susceptibility $\chi$ as a function of the correlation length $\xi$ for various system sizes, signaled by dots with different colors. The black line refers to the relation Eq.~\eqref{chi_corr_relation} obtained from the perturbation theory, with $c_2=67.2$. The inset shows the ratio between $\chi$ and Eq.~\eqref{chi_corr_relation} as a function of $\xi$. The thermodynamic values remain approximately constant. (b) The plot is similar to panel (a) except for the substitution of relation Eq.~\eqref{chi_corr_relation} to \eqref{BKT_relation}, which is the signature of a BKT phase transition. $c_3$ is chosen as 2.6. In the inset, the thermodynamic values of the ratio no longer remain constant.}
    \label{chi_corr}
\end{figure*}

Based on the two-loop result in Eq.~\eqref{2loop}, we can similarly adopt the FSS form of \(\xi\) in Eq.~\eqref{FSS_form} with \(x = \beta-\frac{1}{2\pi}\ln(\beta L)\), that is:
\begin{equation}
    \xi(\beta,L)=L\tilde\xi(\beta-\frac{1}{2\pi}\ln(\beta L)),
\end{equation}
which can ensure the consistency between the FSS ansatz~\eqref{FSS_form} and the thermodynamic behavior in Eq.~\eqref{2loop} in the crossover finite-size window. When \(\xi/L\) is taken as a constant, \(x\) remains constant, and the resulting relationship between \(\beta_L\) and \(L\) is given by:
\begin{equation}
    \beta_L-\frac{1}{2\pi}\ln(\beta_L L) = \text{Const}
\label{relation_beta_L}
\end{equation}
By reformulating Eq.~\eqref{relation_beta_L} and iterating \(\beta_L\), we can express \(\beta_L\) as a function of \(L\):
\begin{equation}
\begin{aligned}
    2\pi\beta_L=&\ln\left(cL\right)+\ln\left(2\pi\beta_L\right)\\
    =&\ln\left(cL\right)+\ln\left[ \ln\left(cL\right)+\ln\left(2\pi\beta_L\right)\right]\\
    =&\cdots,
\end{aligned}
\label{iteration}
\end{equation}
where $c$ refers to a constant different from the $\text{Const}$ in Eq.~\eqref{relation_beta_L}. Thus, through iteration, the results of each order can be obtained:
\begin{equation}
\begin{aligned}
    \beta^{(0)}_L&=\frac{1}{2\pi}\ln\left(cL\right)\\
    \beta^{(1)}_L&=\frac{1}{2\pi}\ln\left(cL\right)+\frac{1}{2\pi}\ln\ln\left(cL\right)\\
    &\cdots.
\end{aligned}
\label{xiL_correction}
\end{equation}
The additional correction terms in higher-order results then lead to a slope slightly larger than \(1/2\pi\) in the large $L$ regime. 

As shown clearly in Fig.~\ref{fig2} (b), the data points agrees with the $\beta_L^{(1)}$ curve for larger $\beta_L$, confirming the presence of the correction term given in Eq.~\eqref{xiL_correction}. For all fixed values of $\xi/L$, the scaling of $\beta_L$ exhibits three distinctive regimes. For $\beta_L < 1.5$, the dependence of $\beta_L$ on $L$ deviates from the logarithmic divergence predicted by Eq.~\eqref{beta_L_diverge}. This corresponds to the initial microscopic regime discussed previously, where the system does not follow the perturbative prediction, and the FSS ansatz no longer holds. For $1.5 < \beta_L < 2.4$, the leading term of the perturbative prediction dominates and the behavior of $\beta_L$ aligns with Eq.~\eqref{beta_L_diverge}. For $\beta_L > 2.4$ up to $\beta_L=4$, the divergence of $\beta_L$ starts to slightly deviate from Eq.~\eqref{beta_L_diverge}, revealing additional correction terms originating from two-loop calculations. 

We thus conclude that $\beta_L$ demonstrates logarithmic divergence with $L$, with further correction terms gradually emerging as $\beta_L$ increases. This divergence precludes a finite-temperature phase transition in the thermodynamic limit. Moreover, we find that higher-order corrections of asymptotic freedom can be directly observed by carefully exploiting the finite-size properties of the system, and $\beta_L$ serves as a good physical quantity to analyze these higher-order corrections.

\section{Susceptibility and Specific Heat}
\label{sec4}

In this section, by examining the relationship between the magnetic susceptibility $\chi$ and the correlation length $\xi$, we provide additional evidence of asymptotic freedom in the 2D Heisenberg model. Then, for completeness, we also discuss the behavior of specific heat $C$ as a function of $\beta$.

As mentioned in Sec.~\ref{sec1}, Patrascioiu et al. \cite{PhysRevB.54.7177} proposed that the system is more likely to undergo a BKT phase transition rather than exhibit asymptotic freedom predicted by the perturbation theory, based on their observation of the relationship between \( \chi \) and \( \xi \). Specifically, for a BKT phase transition, near the critical point, \( \chi \) and \( \xi \) should satisfy the following relation:  
\begin{equation}
    \chi \sim \xi^{1.75}.
    \label{BKT_relation}
\end{equation}
On the other hand, under the two-loop perturbation theory, \( \chi \) and \( \xi \)'s dependence on \( \beta \) is given by \cite{FALCIONI1986671,FALCIONI1985140}:  
\begin{equation}
    \begin{aligned}
    \xi&=c_\xi\frac{e^{2\pi \beta}}{2\pi\beta}\\
    \chi&=c_\chi\frac{e^{4\pi\beta}}{\beta^4}.
    \end{aligned}
\label{2loop_1}
\end{equation}
Thus, Patrascioiu et al. argued that if the system undergoes a BKT phase transition at finite temperature, the ratio \( \chi/\xi^{1.75} \) should remain constant near the transition point. Conversely, if the system exhibits asymptotic freedom, the following relation should always remain valid:
\begin{equation}
    \ln \chi - 2 \ln \xi + 2 \ln \beta = \text{constant}. 
\label{PT_relation}
\end{equation}

They found from numerical data that \( \frac{\chi}{\xi^{1.75}} \) indeed approaches a constant, while the latter expression deviates from a constant, leading them to conclude the presence of a BKT phase transition. Later Allés $et~al.$ observed that \( \frac{\chi}{\xi^{1.75}} \) also does not remain constant in their numerical simulations \cite{ALLES1997677}. However, their results also faced questions due to not using the standard action in their simulation~\cite{patrascioiu1996commentthetwophaseissue}. Furthermore, their observations of \( \chi \) and \( \xi \) still did not support the perturbation theory. Therefore, we re-examine the relationship between \( \chi \) and \( \xi \) in the following analysis to address the problem.

First, it should be noted that the two-loop result in Eq.~\eqref{2loop_1} and the relation in Eq.~\eqref{PT_relation} hold only at sufficiently low temperatures for sufficiently large correlation length. The observed discrepancy between prior numerical simulations and Eq.~\eqref{PT_relation} for finite correlation length does not necessarily imply that the system lacks asymptotic freedom. To better verify the property of asymptotic freedom, we directly examine the relationship between \(\chi\) and \(\xi\) instead of verifying the relation in Eq.~\eqref{PT_relation}. From Eq.~\eqref{2loop_1}, one can obtain \( \beta \) as a function of \( \xi \) at various orders by iteration similar to Eq.~\eqref{iteration}:
\begin{equation}
\begin{aligned}
    2\pi\beta=&\ln\left(\frac{\xi}{ c_\xi}\right)+\ln\left(2\pi\beta\right)\\
    =&\ln\left(\frac{\xi}{ c_\xi}\right)+\ln\left[ \ln\left(\frac{\xi}{ c_\xi}\right)+\ln\left(2\pi\beta\right)\right]\\
    =&\cdots.
\end{aligned}
\label{iteration_1}
\end{equation}
By taking the lowest-order approximation,\( \chi \), as a function of \( \xi \), can be expressed as:  
\begin{equation}
    \chi\sim \xi^2\ln^{-2}\left(\frac{\xi}{c_\xi}\right),
    \label{chi_corr_relation}
\end{equation} 
where $c_\xi$ has already been determined using Bethe ansatz~\cite{HASENFRATZ1990522, HASENFRATZ1990529}: $c_\xi=\frac{e^{1-\pi/2}}{8\times2^{5/2}}$. In Fig.~\ref{chi_corr}, we compare the fit of Eq.~\eqref{chi_corr_relation} and \eqref{BKT_relation} with our data, respectively shown in panel (a) and (b). The flattening data points observed in Fig.~\ref{chi_corr} originate from finite-size effects similar to Fig.~\ref{fig1}, and cannot represent the system's thermodynamic property. Thus, one should focus on the meaningful data points with no flattening tendency. In the main plot, both Eq.~\eqref{chi_corr_relation} and \eqref{BKT_relation} seem to fit the data points reasonably well. However, upon closer inspection, it becomes apparent that for larger values of \( \xi \), Eq.~\eqref{BKT_relation} starts to deviate noticeably from the data points in panel (b). The inset plots show the ratio of \( \chi \) with fits of Eq.~\eqref{chi_corr_relation} and \eqref{BKT_relation} as a function of \( \xi \). The ratio remains approximately constant in panel (a), while in panel (b), it deviates significantly from a constant value. This visually demonstrates that the relationship between \( \chi \) and \( \xi \) in the 2D Heisenberg model does not exhibit the characteristics of a BKT phase transition. Instead, it is well-described by the perturbation theory, which predicts the system's asymptotic freedom.

\begin{figure}[t]
    \centering
    \includegraphics[width=\linewidth]{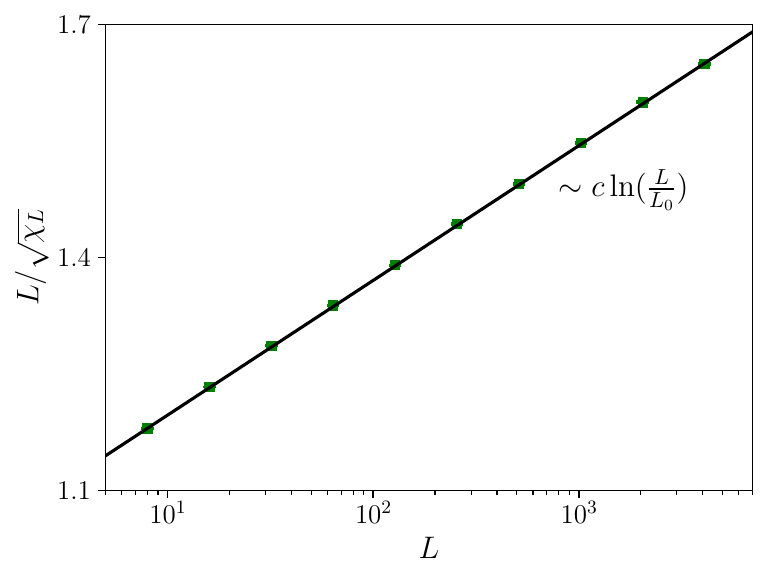}
    \caption{Verification of the proposed relation~\eqref{chi_L_relation}. $L/\sqrt{\chi_L}$ is plotted as a function of $L$. The x-axis is in a logarithmic scale. The black line represents the function $y=c\ln(x/L_0)$ with the parameters chosen as $c=0.0753$ and $L_0=e^{-13.6}$. The excellent logarithmic behavior of data points confirms our proposed relation~\eqref{chi_L_relation}. }
    \label{chi_L}
\end{figure}

In Fig.~\ref{chi_corr}, by examining the scaling of data points with no flattening tendency, we have shown that in the thermodynamic limit, the relationship between $\chi$ and $\xi$ can be well described by Eq.~\eqref{chi_corr_relation}, as predicted by the perturbation theory. For flattening data points, the system is in the FSS regime, where both $\chi$ and $\xi$ are affected by the finite-size effect. We believe that in this regime, $\chi$ exhibits similar behavior as in Eq.~\eqref{chi_corr_relation} with $\xi$ substituted by $L$. Specifically, at the pseudo-critical point $\beta=\beta_L$ where $\xi/L$ is fixed at 1, the value of $\chi(\beta_L)$ is defined as $\chi_L$, which is merely a function of $L$. Since at $\beta=\beta_L$, the correlation length $\xi$ is truncated by the finite system size $L$, by simply substituting $\xi$ by $L$ in Eq.~\eqref{chi_corr_relation}, we expect 
the relationship between $\chi_L$ and $L$ can be similarly expressed as:
\begin{equation}
    \chi_L=aL^2\ln^{-2}(\frac{L}{L_0}),
\label{chi_L_relation}
\end{equation}
with $L_0$ being some constant. Figure~\ref{chi_L} plots $L/\sqrt{\chi_L}$ as a function of $L$ in a semi-log coordinate. The excellent linearity of data points confirms our expected relation in Eq.~\eqref{chi_L_relation}. Therefore, in both the intermediate thermodynamic regime and the FSS regime, the behavior of $\chi$ can be well described by the perturbation theory predictions, providing a robust validation of the theoretical framework.

\begin{figure}
    \centering
    \includegraphics[width=\linewidth]{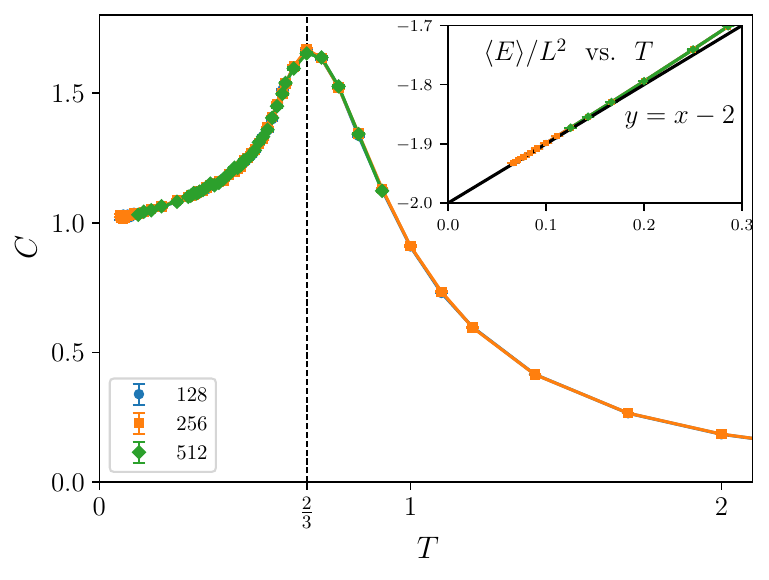}
    \caption{The specific heat $C$ as a function of temperature $T$ for various system sizes. The inset presents the energy density $E/N$ as a function of 
    the temperature. The black line represents the function of $y=x-2$. It is observed that the energy density is proportional to $T$ near $T=0$.}
    \label{heat_capacity}
\end{figure}

Finally, we briefly study the specific heat $C$ to offer additional information on the thermal response of the system.
Figure~\ref{heat_capacity} presents $C$ as a function of the temperature $T$ for different system sizes. The plot demonstrates that the specific heat does not exhibit any divergence as the system size increases, but instead, a smooth peak appears around $T = 2/3$, aligning with half of the mean-field phase transition temperature $T_{\text{mf}}=4/3$. This smooth peak suggests that the specific heat reaches a well-defined maximum, indicating the system’s lack of a sharp phase transition. An interesting observation is that for the 2D Ising and XY models, the peak of the specific heat is also very close to half of the corresponding mean-field transition temperature. 
For the Ising model, the specific heat diverges at \(T_c \approx 2.27 \), while the mean-field transition temperature is \( T_{\text{mf}} = 4 \). In the case of the XY model, the specific heat peak occurs at \( T \approx 1.04 \)~\cite{app11114931}, with the mean-field transition temperature \( T_{\text{mf}} = 2 \). Additionally, it can be observed that as \( N \) increases in the 2D O($N$) model, the peak of the specific heat moves closer to half of the mean-field transition temperature, suggesting that when \( N \) becomes sufficiently large, both values may eventually match. Whether this relationship is a mere coincidence or has a deeper underlying significance may require further investigation. 

Another noteworthy observation is the alignment between the specific heat peak position at $T = \frac{2}{3}$ in Fig.~\ref{heat_capacity} and the crossover point separating the initial microscopic and intermediate thermodynamic regimes in Fig.~\ref{fig1}. This correspondence suggests that the field-theoretical description gradually becomes valid as the system cools to the temperature where heat capacity reaches its maximum. We expect this holds for higher values of $N$. 
In addition to the peak of the \( C-T \) curve in Fig.~\ref{heat_capacity}, the behavior at both ends of the curve, at high and low temperatures, also contains additional information. At high temperatures, since the system's energy has an upper limit due to the constraint on the spin at each lattice site, the heat capacity approaches zero gradually. At low temperatures near zero, the specific heat approaches 1, and the inset further shows the energy density $\langle E\rangle/L^2 \propto T$. This reflects that the 2D Heisenberg model possesses two massless modes near zero temperature.

\section{Conclusion} \label{sec5}
In this work, we perform large-scale MC simulations of the 2D classical Heisenberg model. Through careful and systematic study of the correlation length $\xi$ and the magnetic susceptibility $\chi$, we provide solid evidence to demonstrate that the system satisfies the property predicted by the perturbation theory and has no finite-temperature phase transition. Our main results are listed as follows:
\begin{enumerate}
    \item Without any extrapolation, an evident exponential divergence of the $\xi$ with the inverse temperature $\beta$ is shown in Fig.~\ref{fig1}, including the prefactor $1/2\pi$ in the exponent being observed. Through further extrapolation, our numerical result of $\xi$ coincided with the three-loop perturbative prediction within the error bar when $\beta$ is large enough. 
    \item An FSS formula Eq.~\eqref{xi_FSS_form} for $\xi$ has been proposed, and confirmed in Fig.~\ref{fig2}. Using the formula and the numerical result shown in Fig.~\ref{fig2}, we demonstrate that the pseudo-critical point $\beta_L$ diverges logarithmically as the system size $L$ increases.
    \item The relation~\eqref{chi_corr_relation} of $\chi$ and $\xi$, deriving from the two-loop result in Eq.~\eqref{2loop_1}, is verified numerically in Fig.~\ref{chi_corr}. Moreover, the BKT-like relation in Eq.~\eqref{BKT_relation} between $\chi$ and $\xi$ is ruled out. 
\end{enumerate}
These findings provide robust evidence for the asymptotically-free behavior in the 2D Heisenberg model, suggesting the absence of a BKT phase transition within this system.

As a fundamental model, the 2D classical Heisenberg model has been incessantly studied for decades. While the asymptotically free character of the model has gained well-established consensus, the validity of the field-theoretical prediction and the potential emergence of a BKT phase transition have persisted as subjects of ongoing debate. We revisit this classic and controversial model and provide definitive evidence supporting the asymptotic freedom paradigm while resolving these long-standing ambiguities. Our work sheds light on uncovering some of the non-perturbative aspects of quantum chromodynamics, with broader implications for critical phenomena in condensed matter physics.   
\\
\acknowledgments
We acknowledge the support by the National Natural Science Foundation of China (NSFC) (under Grant No. 12204173, 12275263, and 12274458), National R\&D Program of China (under Grant No. 2023YFA1406500), as well as the Innovation Program for Quantum Science and Technology (under Grant No. 2021ZD0301900). YD is also supported by the Natural Science Foundation of Fujian Province 802 of China (Grant No. 2023J02032).

\appendix

\section*{Appendix A: Details of the Extrapolation} \label{Apen_a}
The fitting equation of $F_\mathcal{\xi}$, as in Ref.~\cite{PhysRevLett.75.1891, PhysRevLett.74.2969}, is typically chosen as the form:
\begin{equation}
    F_\mathcal{\xi}(x)=1+a_1e^{-1/x}+a_2e^{-2/x}+\cdots+a_ne^{-n/x}
\label{extra_fit}
\end{equation}
with $n$ usually no more than 12. Such a fitting requires a substantial amount of data. We systematically perform simulations within the range $1.7 \leq \beta \leq 3$ at intervals of $0.1$ for system sizes $L = 32, 48, 64, 96, 128, 192, 256, 384, 512$. Additionally, simulations for larger system sizes and a few simulations at lower temperatures are conducted, yielding nearly 200 data pairs of $(\beta, L)$ and $(\beta, 2L)$ for fitting.
However, not all data pairs are suitable for fitting. First, due to the limitations of numerical algorithms, the relative error becomes extremely large when $\xi/L$ is too small, making the data unreliable for fitting. Therefore, data points with $\xi / L < 0.1$ are discarded. Second, data points with $\xi / L > 1$ are also excluded due to the truncation of the correlation length by finite system sizes. Furthermore, because of the correction term $O(\xi^{-\omega},L^{-\omega})$ in Eq.~\eqref{extrapolation_eq}, data points with small $\xi$ and $L$ degrade the fitting quality. Consequently, we discard data points with $\xi < 20$, and gradually increase the minimum system size $L_{\mathrm{min}}$ included in the fitting.

During the fitting process, similar to Ref.~\cite{PhysRevLett.75.1891, PhysRevLett.74.2969}, it is observed that finite-size corrections caused by $O(\xi^{-\omega},L^{-\omega})$ are more pronounced in the range $0.3 < \xi / L < 0.7$ than in other intervals. Therefore, two different minimum system sizes $L_{\mathrm{min1}}$ and $L_{\mathrm{min2}}$ are selected for two separate intervals, namely the range $\xi/L\in(0.3,0.7)$ and $\xi/L\in (0,0.3)\cup(0.7,1)$ respectively, with $L_{\mathrm{min1}}\geq L_{\mathrm{min2}}$. $\chi^2/$DF values of different fits are presented in Table~\ref{funcs_fit}, where $\chi^2$ is the sum of squared differences between observed and expected values, normalized by expected values; DF is the degree of freedom in the fit. It can be seen that when $L_{\mathrm{min1}}\geq128$, $L_{\mathrm{min2}}\geq32$, and $n\geq9$, the fitting of Eq.~\eqref{extra_fit} achieves $\chi^2/\text{DF} \leq 1$, indicating relatively reliable results. Apart from $\chi^2/\text{DF}$, an additional approach to test the reliability of the fit is to check the extrapolated values deriving from different system sizes $L$ at the same temperature $\beta$. An adequate fit of $F_\mathcal{O}$ is supposed to ensure the compatibility of these extrapolated values. If a strong discrepancy is observed, to the extent that it distinctly exceeds the estimated errors, this may indicate that the fitting of $F_\mathcal{O}$ is not adequate and requires further improvement by increasing $n$ and $L_{\mathrm{min}}$. On the other hand, given a suitable fit for $F_\mathcal{O}$, combining the extrapolated results obtained from data at different system sizes for a given temperature can also effectively reduce the error bar of the final extrapolated result.

In the main text, for convenience, only the optimal fit of $n=9$ and $(L_{\mathrm{min1}}=128, L_{\mathrm{min2}}=64)$ is used. This is based on the observation in Table~\ref{differ_funcs} that the systematic errors due to different fits of $F_\xi$ are generally within the statistical error range. Here, we present more details of the optimal fit. Table~\ref{fit_detail} provides the specific fitting parameters obtained for $n = 9$ with different values of $(L_{\mathrm{min1}},L_{\mathrm{min2}})$. The extrapolated correlation lengths at different temperatures and system sizes using the optimal fit are shown in Table~\ref{differ_beta}. For most temperatures, the extrapolated values obtained for various system sizes remain within the statistical error range, which further demonstrates the reliability of the extrapolation method and the fitted $F_\xi$.

\begin{table*}[!ht]
    \centering
    \caption{The $\chi^2/$DF values of the fit of $F_\xi$ based on Eq.~\eqref{extra_fit} with different $n$ and $(L_{\mathrm{min1}},L_{\mathrm{min2}})$.}
    \label{funcs_fit}
    \begin{tabularx}{\textwidth}{XXXXXXX}
    \hline\hline
        $L_\mathrm{min}$ & n=7 & n=8 & n=9 & n=10 & n=11 & n=12 \\ \hline
        (96, 32) & 340.4/150 & 232.7/149 & 179.4/148 & 179.3/147 & 173.7/146 & 170.7/145 \\ 
        (96, 64) & 319.2/134 & 216.6/133 & 162.4/132 & 162.4/131 & 155.9/130 & 153.5/129 \\ 
        (128, 32) & 266.5/139 & 174.5/138 & 125.9/137 & 125.8/136 & 119.8/135 & 116.3/134 \\ 
        (128, 64) & 245.2/123 & 158.5/122 & 109.0/121 & 109.0/120 & 101.9/119 & ~~99.1/118 \\ 
        (128, 128) & 232.5/109 & 147.9/108 & 100.8/107 & 100.8/106 & ~~92.4/105 & ~~90.1/104 \\ 
        (192, 32) & 235.0/128 & 151.9/127 & 113.4/126 & 113.3/125 & 109.1/124 & 106.2/123 \\ 
        (192, 64) & 212.3/112 & 135.7/111 & ~~96.5/110 & ~~96.5/109 & ~~91.5/108 & ~~89.1/107 \\ 
        (192, 128) & 198.2/98 & 125.8/97 & ~~88.4/96 & ~~88.3/95 & ~~82.2/94 & ~~80.1/93 \\ \hline\hline
    \end{tabularx}
\end{table*}

\begin{table*}[!ht]
    \centering
    \caption{The extrapolated values of $\xi$ using different fits of $F_\xi$, for $\beta=2.4, 2.6, \cdots, 3.8$. The three numbers in the brackets of the first column represent the values of \( n \), \( L_{\text{min1}} \), and \( L_{\text{min2}} \) when fitting \( F_\xi \) in Eq.~\eqref{extra_fit}. The values in each column need to be multiplied by the factor in the first line's bracket.}
    \label{differ_funcs}
    \begin{tabularx}{\textwidth}{XXXXXXXXX}
    \hline\hline
        $F_\xi$ & 2.4($\times10^3$) & 2.6($\times 10^3$) & 2.8($\times 10^4$) & 3($\times10^4$) & 3.2($\times 10^5$) & 3.4($10^6$) & 3.6($\times10^6$) & 3.8($\times 10^7$) \\ \hline
        (9,128,64) & 2.6138(61) & 8.616(60) & 2.875(19) & 9.636(79) & 3.199(31) & 1.064(12) & 3.594(58) & 1.163(31) \\ 
        (9,128,128) & 2.6134(63) & 8.620(61) & 2.883(19) & 9.691(78) & 3.220(29) & 1.072(12) & 3.616(56) & 1.169(33) \\ 
        (10,128,64) & 2.6135(63) & 8.621(58) & 2.873(18) & 9.634(75) & 3.193(31) & 1.061(12) & 3.597(54) & 1.165(31) \\ 
        (10,128,128) & 2.6145(62) & 8.619(60) & 2.881(19) & 9.678(77) & 3.219(30) & 1.070(12) & 3.627(60) & 1.173(30) \\ 
        (11,128,64) & 2.6105(62) & 8.612(58) & 2.880(18) & 9.631(77) & 3.187(30) & 1.062(12) & 3.586(53) & 1.186(27) \\ 
        (11,128,128) & 2.6122(65) & 8.646(54) & 2.892(19) & 9.682(81) & 3.211(30) & 1.070(12) & 3.613(60) & 1.195(27) \\ \hline\hline
    \end{tabularx}
\end{table*}

\begin{table*}[!ht]
    \centering
    \caption{Specific values of parameters obtained from the fit of Eq.~\eqref{extra_fit} for $n=9$ and different $(L_{\mathrm{min1}},L_{\mathrm{min2}})$.}
    \label{fit_detail}
    \begin{tabularx}{\textwidth}{XXXXXXXXXXX}
    \hline\hline
$L_{\text{min}}$ & $a_1(\times10^{-1})$ & $a_2(\times10^1)$ & $a_3(\times10^3)$ & $a_4(\times10^4)$ & $a_5(\times10^5)$ & $a_6(\times10^6)$ & $a_7(\times10^6)$ & $a_8(\times10^6)$ & $a_9(\times10^6)$ & $\chi^2/$DF \\ \hline
        (96, 32) & 7.5(21) & 7.9(14) & -2.08(36) & 3.37(48) & -2.84(37) & 1.31(17) & -3.37(44) & 4.60(64) & -2.59(39) & 180.3/152 \\ 
        (96, 64) & 7.5(22) & 8.0(14) & -2.10(38) & 3.40(50) & -2.86(39) & 1.32(18) & -3.41(47) & 4.66(69) & -2.63(42) & 161.9/129 \\ 
        (96, 128) & 7.5(23) & 8.0(16) & -2.09(41) & 3.39(56) & -2.86(43) & 1.32(20) & -3.40(54) & 4.65(79) & -2.62(49) & 152.5/109 \\ 
        (128, 32) & 7.6(19) & 7.9(12) & -2.06(33) & 3.36(44) & -2.83(34) & 1.31(15) & -3.39(41) & 4.64(59) & -2.62(36) & 126.9/141 \\  
        (128, 64) & 7.5(19) & 7.9(13) & -2.07(34) & 3.38(46) & -2.86(35) & 1.32(16) & -3.43(43) & 4.69(62) & -2.65(38) & 108.5/118 \\  
        (128, 128) & 7.6(20) & 7.9(14) & -2.06(37) & 3.37(50) & -2.85(39) & 1.32(18) & -3.41(49) & 4.68(71) & -2.64(44) & ~~99.1/98 \\ 
        \hline
    \end{tabularx}
\end{table*}

\begin{table*}[!ht]
    \centering
    \caption{The extrapolated values of $\xi$ at different temperatures and system sizes using specific fit of $F_\xi(x)$ for $n=9$, $(L_\text{min1}=128,L_\text{min2}=64)$. The ``mean" column combines 
    extrapolated results for different system sizes. The ``3-loop" column shows values obtained from the three-loop perturbative result in Eq.~\eqref{3loop}. Values in each line need to be multiplied by the factors in the last column.}
    \label{differ_beta}
    \begin{tabularx}{\textwidth}{XXXXXXXXXXX}
    \hline\hline
        beta$\backslash$L & 64 & 128 & 256 & 512 & 1024 & 2048 & 4096 & mean & 3-loop & factor \\ \hline
        1.6  & 1.905(3) & 1.903(4) & 1.903(8) & 1.90(3) & - & - & - & 1.903(3) & 2.71799 & $\times10^1$ \\ 
        1.7 & 3.465(9) & 3.458(7) & 3.452(6) & 3.45(2) & - & - & - & 3.455(4) & 4.81207 & $\times10^1$ \\ 
        1.8 & 6.55(3) & 6.47(1) & 6.46(1) & 6.48(1) & 6.47(4) & - & - & 6.467(7) & 8.54567 & $\times10^1$ \\ 
        1.9 & 1.230(7) & 1.220(6) & 1.223(3) & 1.223(2) & 1.223(3) & 1.231(7) & - & 1.223(1) & 1.52179 & $\times10^2$ \\ 
        2.0 & 2.34(2) & 2.28(1) & 2.288(8) & 2.283(5) & 2.293(4) & 2.290(6) & - & 2.289(3) & 2.71673 & $\times10^2$ \\ 
        2.1 & 4.25(4) & 4.27(3) & - & 4.23(2) & 4.227(8) & 4.229(6) & 4.24(2) & 4.230(5) & 4.86090 & $\times 10^2$ \\ 
        2.2 & 7.84(8) & 7.93(7) & 7.81(6) & 7.77(4) & 7.78(2) & 7.81(1) & 7.79(1) & 7.798(8) & 8.71528 & $\times10^2$ \\ 
        2.3 & 1.41(2) & 1.43(2) & 1.44(1) & 1.45(1) & 1.429(7) & - & 1.425(3) & 1.428(2) & 1.56555 & $\times10^3$ \\ 
        2.4 & 2.61(4) & 2.62(4) & 2.64(3) & 2.63(2) & 2.64(2) & 2.60(1) & 2.610(9) & 2.613(6) & 2.81710 & $\times10^3$ \\ 
        2.5 & 4.81(9) & 4.77(7) & 4.67(6) & 4.73(5) & - & 4.71(3) & 4.77(3) & 4.74(2) &  5.07730 & $\times10^3$ \\ 
        2.6 & 8.7(1) & - & 8.5(1) & 8.7(1) & 8.59(8) & - & - & 8.62(6) & 9.16443 & $\times 10^3$ \\ 
        2.7 & 1.60(3) & 1.59(2) & 1.59(3) & 1.59(2) & 1.59(2) & - & - & 1.59(1) & 1.65643 & $\times10^4$ \\ 
        2.8 & 2.97(7) & 2.89(6) & 2.90(5) & 2.86(4) & 2.88(4) & 2.87(4) & - & 2.88(2) & 2.99775 & $\times10^4$ \\ 
        2.9 & 5.3(1) & 5.4(1) & 5.3(1) & 5.43(9) & 5.29(8) & 5.24(7) & - & 5.31(4) & 5.43166 & $\times 10^4$ \\ 
        3.0 & 9.5(2) & 9.6(2) & 9.7(2) & 9.7(2) & 9.6(2) & 9.6(2) & - & 9.65(8) & 9.85266 & $\times10^4$ \\ 
        3.1 & 1.72(5) & 1.72(4) & 1.77(4) & 1.75(3) & 1.76(3) & 1.76(3) & - & 1.75(1) & 1.78907 & $\times10^5$ \\ 
        3.2 & - & 3.14(9) & 3.21(8) & 3.21(7) & 3.13(6) & 3.31(5) & - & 3.22(3) & 3.25180 & $\times 10^5$ \\ 
        3.3 & - & - & 6.0(1) & 6.0(1) & 5.8(1) & 5.8(1) & 5.9(1) & 5.89(5) & 5.91589 & $\times10^5$ \\ 
        3.4 & - & - & - & 1.04(3) & 1.09(3) & 1.05(2) & 1.07(2) & 1.06(1) & 1.07719 & $\times10^6$ \\ 
        3.5 & - & - & - & - & 1.96(5) & 1.93(5) & 1.92(4) & 1.93(3) & 1.96300 & $\times 10^6$ \\ 
        3.6 & - & - & - & - & - & 3.63(9) & 3.56(8) & 3.59(6) & 3.58000 & $\times 10^6$ \\ 
        3.7 & - & - & - & - & - & - & 6.6(1) & 6.6(1) & 6.53375 & $\times10^6$ \\ 
        3.8 & - & - & - & - & - & - & 1.17(3) & 1.17(3) & 1.19329 & $\times10^7$ \\ \hline\hline
    \end{tabularx}
\end{table*}

\bibliographystyle{apsrev4-1}
\bibliography{short_range_heisenberg, SR_Heisenberg}

\end{document}